\documentclass{aa}
\usepackage{epsfig,graphicx,times}
\newcommand{\kms}{{{km~s}$^{-1}$}}
\newcommand{\teff}{{$T_\mathrm{eff}$}}

\newcommand{\vt}{{$v_\mathrm{t}$}}

\begin{document}
\title{A Be star with a low nitrogen abundance in the 
SMC cluster NGC\,330}
\subtitle{}

\author{D.J. Lennon\inst{1}
\and J.-K. Lee\inst{2}
\and P.L. Dufton\inst{2}
\and R.S.I. Ryans\inst{2}} 

\offprints{D.J. Lennon,\email{djl@ing.iac.es}}

\institute{The Isaac Newton Group of Telescopes, 
Apartado de Correos 321, E-38700
Santa Cruz de La Palma, Canary Islands, Spain
\and Department of Pure \& Applied Physics, The Queen's University 
           of Belfast, BT7 1NN, Northern Ireland, UK}

\date{Received: 2004 / Accepted}

\abstract{
High-resolution UVES/VLT spectra of B\,12, an extreme pole-on Be star in the SMC
cluster NGC\,330, have been analysed using non-LTE model atmospheres to obtain
its chemical composition relative to the SMC standard star
AV\,304. We find a general underabundance of metals which can be understood in
terms of an extra contribution to the stellar continuum due to emission from a
disk which we estimate to be at the $\sim$25\% level.  When this is corrected
for, the nitrogen abundance for B\,12 shows no evidence of 
enhancement by rotational mixing as has been found in other non-Be B-type 
stars in NGC\,330, and is {\em inconsistent} with evolutionary models which
include the effects of rotational mixing. A second Be star, NGC\,330-B\,17,
is also shown to have no detectable nitrogen lines. Possible
explanations for the lack of rotational mixing in these rapidly
rotating stars are discussed, one promising solution being the
possibility that magnetic fields might inhibit rotational mixing.

}

\titlerunning{A Be star with a low nitrogen abundance in NGC\,330}

\maketitle

\section{Introduction}

The open cluster NGC\,330 has become one of the most studied young clusters in
the Small Magellanic Cloud (SMC).  There have been many spectroscopic studies
of individual massive stars in the cluster, attention being concentrated on the
most massive and luminous giants and supergiants (Lennon et al.\ \cite{Len03b}, 
\cite{Len96}, \cite{Len93}; 
Hill et al.\cite{Hil97}; Hill \cite{Hil99}; Spite et al.\cite{Spi91}).  The most salient and
interesting result from these studies is that all the B-type giants and
supergiants are strongly nitrogen enriched.
This striking result echoes the findings of Venn
(\cite{Ven99}), Dufton et al.\ (\cite{Duf00}) and Trundle at al.\ (\cite{Tru04}) 
that most early-type supergiants in the SMC are enriched in nitrogen by around
a factor of 10.  It is also interesting to note that similarly  large nitrogen
enrichments have also been found for many, but not all, O-type stars in the SMC
(Bouret et al.\ \cite{Bou03}, Hillier et al.\ \cite{Hil03}).

One of the most favoured explanations as to the cause of these enrichments is
that rotationally induced mixing is responsible for polluting the atmospheres
of these stars with CN processed material (Maeder \& Meynet \cite{Mae01}, Heger
\& Langer \cite{Heg00}). In the context of stellar rotation, the cluster NGC\,330 is
again prominent  since it has the highest known fraction of Be stars of any
known cluster (Feast \cite{Fea72}, Grebel et al.\ \cite{Gre96}, Mazzali et al.\
\cite{Maz96}, Keller \& Bessell \cite{Kel98}). Indeed it has been suggested
that this high incidence of Be stars might be linked to the low metallicity of
the cluster (Maeder et al.\ \cite{Mae99}). This apparent dependence of the
frequency of Be stars on metallicity has led to suggestions
that stellar rotational rates might increase with decreasing metallicity, and
hence increasing redshift, with far-reaching consequences for stellar
evolution.  For example, Meynet \& Maeder (\cite{Mey03}) have suggested that
stellar rotation might play a key role in the production of nitrogen in the
Universe.  

However, as discussed by Lennon (\cite{Len03a}), Herrero \& Lennon
(\cite{Her04}) and Walborn \& Lennon (\cite{Wal04}), it is difficult to
reconcile the predictions of various theories with the observed distributions
of stellar rotational velocities, or the observed distribution of massive stars
in the HR diagram.  Lennon (\cite{Len03a}) discussed the case of NGC\,330 and
it seems clear that, even allowing for inclination effects, the rotational
velocities of the nitrogen enriched B-type giants are much lower than those
needed to produce  enrichment in the models.   
A major concern for any inferences
drawn about the role of rotation on massive  star evolution is that there is a
very strong observational bias towards low-$v$\,sin\,$i$ stars, for 
the simple reason that
spectra with even moderate rotational velocities are extremely difficult to
analyse.  Furthermore, from studies of the distribution of rotational
velocities of massive stars (Wolff \& Simon \cite{Wol97}, Day \& Warner
\cite{Day75}) it seems reasonable to assume that the majority of the
low-$v$\,sin\,$i$ stars are indeed slow rotators. Ideally one needs to consider
pole-on stars which are known to have high rotational
velocities. However deducing rotational velocities from the spectra of
narrow-lined normal B-type stars is difficult, whereas 
pole-on Be stars fit this description extremely well. They are
acknowledged to be fast rotators and the orientation angle of the rotation
axis may be deduced from the morphology of the disk emission lines, rather
than just the widths of absorption lines.   

Although the probability of observing an extreme pole-on star
is low, the very high fraction of Be stars in the cluster NGC\,330 makes
this more likely. Furthermore Baade et al.\
(\cite{Baa02}) have coincidentally carried out a search for spectral
variability with UVES/VLT in two low-$v$\,sin\,$i$ Be stars in the cluster, one
of which, B\,12 (Robertson \cite{Rob74}), is almost pole-on with a
$v$\,sin\,$i$ of only 40 km/s.  Lennon et al.\ (\cite{Len93}) have also noted
that this Be star had a low $v$\,sin\,$i$, as had Hummel et al.\ (\cite{Hum01})
in a FORS/VLT study of Be stars in the cluster.  In
this paper we present a description of the high resolution spectrum of B\,12
obtained by Baade et al., and attempt a detailed analysis of the abundances
using non-LTE model atmospheres, paying special attention to the nitrogen 
lines in its spectrum.

\section{Observational data}

\begin{figure}\centering
 \hspace*{-5mm}\psfig{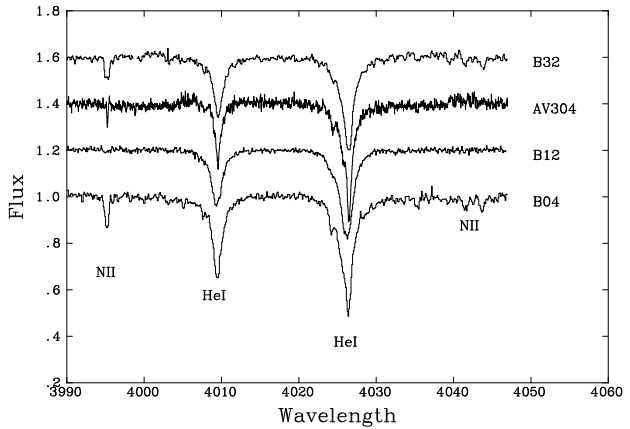}\\

 \caption{A comparison of the 3990--4050 \AA\ regions for 
the nitrogen rich stars B\,32 and B\,04 in NGC\,330, the
nitrogen normal star AV304, and the Be star B\,12. We indicate the presence 
of the He\,{\sc i} lines at 4009 and 4026 \AA , together with the 
N\,{\sc ii} lines at 3995, 4041 and 4043 \AA . Note that 
the extraordinary sharpness of the lines in AV304 permits the 
secure identification on the N\,{\sc ii} line at 3995 \AA\ while the 
high S/N ratio of the data for B12 permits a tentative detection of this
same line, which is significantly weaker than in the other NGC\,330
stars B\,04 and B\,32. }\label{fig:comparison} 
\end{figure}

The data for B12, 
which were kindly made available to us by Thomas Rivinius, are
described in detail by Baade et al.\ (\cite{Baa02}), and consist of 23
individual observations of B\,12 using UV-Visual Echelle Spectrograph (UVES) on
the VLT.  All 23 spectra, each of which had a S/N ratio of $\approx$ 80, 
were coadded as Baade et al.\ found little evidence for variability.  It was
immediately clear that the spectrum of B\,12 is peculiar compared to those of
the other B-type stars in NGC\,330 observed 
by Lennon et al.\ (\cite{Len03b}) in that the N\,{\sc ii} spectrum
was extremely weak.  This is illustrated in Fig.~\ref{fig:comparison}
where we compare the spectral region containing the N\,{\sc ii} lines at 
3995, 4041 and 4043 \AA\
with that of other SMC B-type stars; 
NGC\,330-B\,32 (also observed by Baade et al),
NGC\,330-B\,04 (observed with UVES/VLT by DJL), and the standard AV\,304 
(observed with UVES/VLT by Rolleston et al.\ (\cite{Rol03})).  
It is clear from this figure that the \ion{N}{ii} lines are at the limit
of detectability in the spectrum of B\,12 despite its very high S/N ratio.  
The other B-type stars in NGC\,330 are
in sharp contrast to B\,12, but note that Lennon et al.\ (\cite{Len03b}) found these
objects to be nitrogen rich with respect to the standard AV\,304 by
approximately a factor of 10.   
This is a surprising result as it implies that nitrogen abundance of
B\,12 must be very low, as can be seen by a comparison with the spectrum of
AV\,304 (note that the N\,{\sc ii} line at 3995\AA\ is only visible in this
star because of its very low projected rotational velocity).  

While absorption lines of N\,{\sc ii} are mostly absent from the spectrum
of B\,12, many other metal absorption lines typical of an early
B-type star {\em are} present; in particular there are 
absorption lines due to O\,{\sc ii},
Si\,{\sc iii}, Mg\,{\sc ii} and C\,{\sc ii}. 
We also note the presence of several emission lines due to 
allowed transitions of Fe\,{\sc ii}
and Si\,{\sc ii}.
The equivalent widths (EWs) of the 
metal absorption lines were measured using a {\sc starlink}
spectral analysis program DIPSO (Howarth et al.\ \cite{How96}). Non-linear 
least square fitting routines  were used with 
Gaussian profiles being adopted for the absorption lines and
low order polynomials for the adjacent continuum regions. This 
procedure also yielded an estimate of the radial
velocity of 152 $\pm$ 5 \kms. For the hydrogen lines,
the profiles were fitted with the
continuum levels being defined at $\pm$16 \AA\ from the line
centre. The equivalent widths are listed in Table~\ref{tab:b12ew}.
Generally the lines with EWs $\geq$ 15 m\AA\ were well observed, with
convincing fits being obtained for several weaker lines. 

\begin{table}
\caption{Equivalent widths (m\AA) for B\,12. Typical measurement uncertainties
is $<$5 m\AA.}\label{tab:b12ew}
\centering
\begin{tabular}{lrllrllr}
\hline\hline\noalign{\smallskip}
Line & EWs && Line & EWs && Line & EWs\\
\noalign{\smallskip}\hline\noalign{\smallskip}
\multicolumn{2}{l}{C\,{\sc ii}} && \multicolumn{2}{l}{O\,{\sc ii}} 
                                                      && \multicolumn{2}{l}{Si\,{\sc ii}} \\
3918.98 & 55                    && 4132.81 & 13       && 4128.07 & $\leq$10		  \\
3920.68 & 56                    && 4253.90 & 28       && 4130.89 & $\leq$10		  \\
4074.52 & $\leq$10              && 4317.14 & 22       &&	      & 		  \\
4267.02 & 91                    && 4319.63 & 22       && \multicolumn{2}{l}{Si\,{\sc iii}}\\
        &                       && 4349.43 & 22       && 4552.38 & 59			  \\
\multicolumn{2}{l}{N\,{\sc ii}} && 4395.94 & 11       && 4567.82 & 50			  \\
3995.00 & 12                    && 4414.90 & 35       && 4574.86 & 38			  \\
4601.48 & $\leq$10              && 4416.97 & 7:       &&	    &			  \\
4607.16 & $\leq$10              && 4452.37 & 9:       && \multicolumn{2}{l}{Si\,{\sc iv}} \\
4621.29 & $\leq$10              && 4590.97 & 26       && 4088.85 & 10			  \\
4643.09 & $\leq$10              && 4595.96 & 16       && 4116.10 & $\leq$10		  \\
        &                       && 4609.44 & 6:       &&	 &			  \\
\multicolumn{2}{l}{O\,{\sc ii}} && 4613.11 & $\leq$10 && \multicolumn{2}{l}{Fe\,{\sc ii}} \\
3954.37 & 16                    && 4638.86 & 21       && 4395.78 & 12			  \\
4069.62 & 35                    && 4641.82 & 35       &&	 &			  \\
4072.16 & 25                    && 4649.14 & 37       && \multicolumn{2}{l}{Mg\,{\sc ii}} \\
4075.85 & 28                    && 4650.84 & 13       && 4481.13 & 43			  \\
4078.84 & $\leq$10              && 4661.63 & 13       &&	 &			  \\
4083.91 & $\leq$10              && 4676.24 & 13       && \multicolumn{2}{l}{Al\,{\sc ii}} \\
4084.65 & 9                     && 4699.00 & 9:       && 4528.91 & 13:  		  \\
4087.16 & $\leq$10              && 4705.35 & 11       &&         &                        \\
4089.29 & 10                    &&         &          &&         &                        \\
\noalign{\smallskip}\hline\noalign{\smallskip}				       
\end{tabular}
\end{table}

\begin{table*}[t]
\caption{Atmospheric parameters and abundances of AV\,304 together with the 
differential abundances derived for B\,12 with the number of lines considered
being given in parentheses. The errors for the former are taken from
Hunter et al.\ (\cite{Hun04}). For the latter
they are the standard error among the lines of the same 
element; no error is given for N\,{\sc ii} and Mg\,{\sc ii} since only
one feature contributed to these abundances.  Also listed in this table
are the parameters and abundances which one would obtain for B\,12 assuming
a disk was contributing 25\% to the continuum flux.  There are two sets of 
stellar parameters for each disk/no disk model; in each case the higher effective temperature
is the one derived by taking the mean of the upper and lower limits set by the absence of the
Si\,{\sc iv} and Si\,{\sc ii} lines, while the lower value represents the effective
temperature set by the absence of the Si\,{\sc ii} lines only. 
}\label{tab:abs_abn}
\begin{tabular}{clrllllr}
\hline\hline\noalign{\smallskip}
 & AV\,304$^*$ & & \multicolumn{4}{c}{B\,12}  & \\ 
              &          &        & no disk & no disk   & 25\% disk & 25\% disk &   \\
\noalign{\smallskip}\hline\noalign{\smallskip}
\teff\/ (K)      & 27\,500         &     &~23\,000         & ~21\,500         &  ~24\,000 & ~22\,500 &          \\
log~$g$ (cm s$^{-2}$)      & 3.90            &     &~3.60            & ~3.45	         &  ~4.00 & ~3.85 &           \\ 
\vt  (\kms)         & 3 && ~3 & ~3 & ~3 & ~3 \\ \noalign{\smallskip}
 ~C\,{\sc ii} & 7.36 $\pm$ 0.12 & ( 3)&$-0.06$ $\pm$ 0.02 & $-0.15$ $\pm$ 0.04 & ~0.19 $\pm$ 0.02& ~0.09 $\pm$ 0.04& ( 2)    \\ 
 ~N\,{\sc ii} & 6.55            & ( 1)&$-0.35$		  & $-0.33$  	    & $-0.19$ & $-0.17$ & ( 1)    \\
 ~O\,{\sc ii} & 8.13 $\pm$ 0.10 & (42)&$-0.62$ $\pm$ 0.13 & $-0.43$ $\pm$ 0.14 & $-0.35$ $\pm$ 0.12 & $-0.19$ $\pm$ 0.10& (13)    \\
 ~Mg\,{\sc ii}~ & 6.77          & ( 1)&$-0.33$		  & $-0.44$  	    & $-0.14$ & $-0.24$ & ( 1)      \\
 ~Si\,{\sc iii}& 6.76 $\pm$ 0.18& ( 4)&$-0.43$ $\pm$ 0.12 & $-0.27$ $\pm$ 0.12 & $-0.05$ $\pm$ 0.10& ~0.08 $\pm$ 0.09& ( 3)    \\
\noalign{\smallskip}\hline\noalign{\smallskip}
\end{tabular}\\
$*$ Hunter et al.\ (\cite{Hun04}). Rolleston et al.\ (\cite{Rol03}) derived 
\teff\/ = 27\,500 K, log~$g$ = 3.8, \vt = 5 \kms\ in their LTE analysis.\\
\end{table*}

\section{Method of analysis}\label{analysis}

We have already referred to the marked weakness of the nitrogen
lines in B\,12 compared to other nitrogen rich B-type giants in
NGC\,330.  Since it is generally accepted that Be stars are
B-type stars near the end of the main-sequence it is especially
interesting to try to quantify the atmospheric nitrogen abundance
of B\,12 in order to see to what degree, if any, the nitrogen content
has been enhanced by rotational mixing.  In the following subsections
we present the results of a non-LTE model atmosphere analysis of B\,12 relative
to a B-type SMC standard.  Clearly the use of plane-parallel
models for a star like B12 which is a fast rotator with a disk is
questionable. However we consider it a reasonable first approximation
and return to the question of its reliability in the subsequent discussion.

\subsection{Model atmosphere calculations} 

The analysis is based on grids of non-LTE model atmospheres calculated using
the codes {\sc tlusty} and {\sc synspec} (Hubeny \cite{Hub88}; 
Hubeny \& Lanz \cite{Hub95};
Hubeny et al.\ \cite{Hub98}).  Details  of the methods can be found in Ryans et
al.\ (\cite{Rya03}), while the grids are discussed in more detail by Dufton et
al.\ (\cite{Duf04}) and Lee et al.\ (\cite{Lee04}). 

Briefly we used a grid appropriate to the SMC metallicity that has been
generated with an iron abundance of $[\frac{Fe}{H}] =$ 6.8, i.e.\ 0.7 dex
less than Galactic. The model grid covers a range of effective temperatures
from 12\,000 to 35\,000~K, logarithmic gravities from 4.5 dex down to close to
the Eddington limit and microturbulences from 0 to 30 km~s$^{-1}$. For any set
of atmospheric parameters, five models were then calculated keeping the iron
abundance fixed but allowing the light element (e.g.\ C, N, O, Mg and Si)
abundances  to vary from +0.8 dex to $-$0.8 dex around their base values.  
These models are then used to calculate spectra, which provide  theoretical  
hydrogen and helium line profiles and equivalent widths for light metals 
for a range of  abundances. The theoretical equivalent widths were measured 
using IDL programs to integrate the flux within a certain wavelength range 
chosen to include the line of interest.

\subsection{Comparison star}

Hunter et al.\ (\cite{Hun04}) have analysed UVES/VLT spectroscopy (Rolleston
et al.\ \cite{Rol03}) of an SMC main sequence star, AV\,304. They used the same  
grid of TLUSTY model atmospheres as discussed above, together with
similar techniques to those adopted here to estimate the atmospheric
parameters and element abundances (see Sect. \ref{atm} and \ref{Results}). As
such, these results provide a useful comparison to those for B12 and
we summarize the results of their analysis in Table \ref{tab:abs_abn}.


\subsection{Estimation of atmospheric parameters} \label{atm}

The atmospheric parameters, surface gravity and effective temperature, 
are inter-related, and their derivation requires an iterative approach.
We adopted an appropriate value of the surface gravity and then
estimated the effective temperature using the silicon ionization balance.
This temperature estimate was then used to deduce a new value for the 
surface gravity from fitting the Balmer lines, and the process iterated 
to convergence. Unfortunately in the spectra of B\,12, it was not possible
to distinquish either Si\,{\sc ii} or Si\,{\sc iv} lines. However their absence
effectively provided constraints on the lower and upper limits of the effective
temperature.  The mean of the derived lower and upper limits, which differ by
approximately 2\,500~K, was then taken to be the stellar effective temperature, 
although we will discuss below the effect of using these upper and lower 
bounds on the derived abundances. There were insufficient metal lines to 
determine the microturbulent velocity and we have therefore adopted the value
of Hunter et al.\ (\cite{Hun04}) of \vt\/ = 3 \kms. We note that due to the weakness 
of the metal lines in the spectrum of B\,12, the abundance estimates are
insensitive to the value adopted for the microturbulence.

\subsection{Abundance estimates}
\label{Results} 

The metal lines were used, along with the atmospheric parameters derived
above, to deduce absolute abundances for B\,12. A typical observational 
uncertainty of $\leq$5 m\AA\ in the EW estimate of an individual line would 
result in an abundance error less than 0.1 dex. Normally only metal lines with
an EW of greater than 15 m\AA\ were considered. However for the ion \ion{N}{ii}, the
only feature observed was at 3995\AA\ and had an equivalent width of 12 m\AA.
We have therefore included this line but note that the corresponding
nitrogen abundance should be considered as an upper limit. A detailed 
line-by-line differential analysis (summarised in 
Table~\ref{tab:abs_abn}) of B\,12  was then carried out relative to AV\,304, as 
this should minimise the effect of systematic errors.

\section{Discussion}       

The most important and obvious result of this investigation is that B\,12 
does {\em not} exhibit any evidence for the kind of nitrogen enrichments 
found in other evolved B-type stars in NGC\,330.  
Surprisingly however {\em all} elements in B\,12 
show a general under-abundance relative to AV\,304
and other NGC\,330 stars (Lennon et al.\ \cite{Len03b}). For example
compared with AV\,304, the under-abundances range from $-$0.06 dex for 
C to $-$0.62 dex for O, with N, Mg and Si showing rather similar
under-abundances of $-0.35$ to $-0.45$ dex.

Of course our
estimate of the effective temperature is derived by simply taking the mean
of the upper and lower bounds implied by the absence of Si\,{\sc iv} and 
Si\,{\sc ii} lines.  While the effective temperature could lie anywhere
within this range, increasing the effective temperature (and, for consistency,
the surface gravity) only increases the discrepancy between B\,12 and AV\,304.
Decreasing our effective temperature to its lower limit of 21\,500\,K does improve
the overall situation for all elements (see Table~\ref{tab:abs_abn}) but still leaves us
with significant systemic differences with AV\,304.  There are two
likely causes of these differences which are suggested by the Be nature
of B\,12; veiling of the continuum due to a disk, and departures from 
standard non-rotating plane parallel models caused by rapid rotation.

 \begin{figure}\centering
 \begin{tabular}{l}
 \hspace*{-5mm}\psfig{file=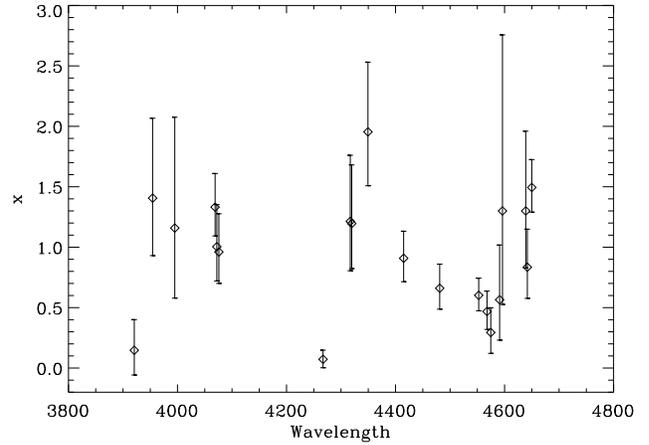,width=0.5\textwidth}\\
 \end{tabular} 
  \caption{The fractional contribution, x, of the disk relative to the stellar 
  continuum which is required for metal lines in B\,12 to have the same
  abundance as in the comparison star AV\,304. These calculations have been performed
  for the \teff = 23\,000, log\,g = 3.60 model from Table~\ref{tab:abs_abn}.
}\label{fig:scale}
 \end{figure}

 \begin{figure*}\centering
  \hspace*{-8mm} \psfig{file=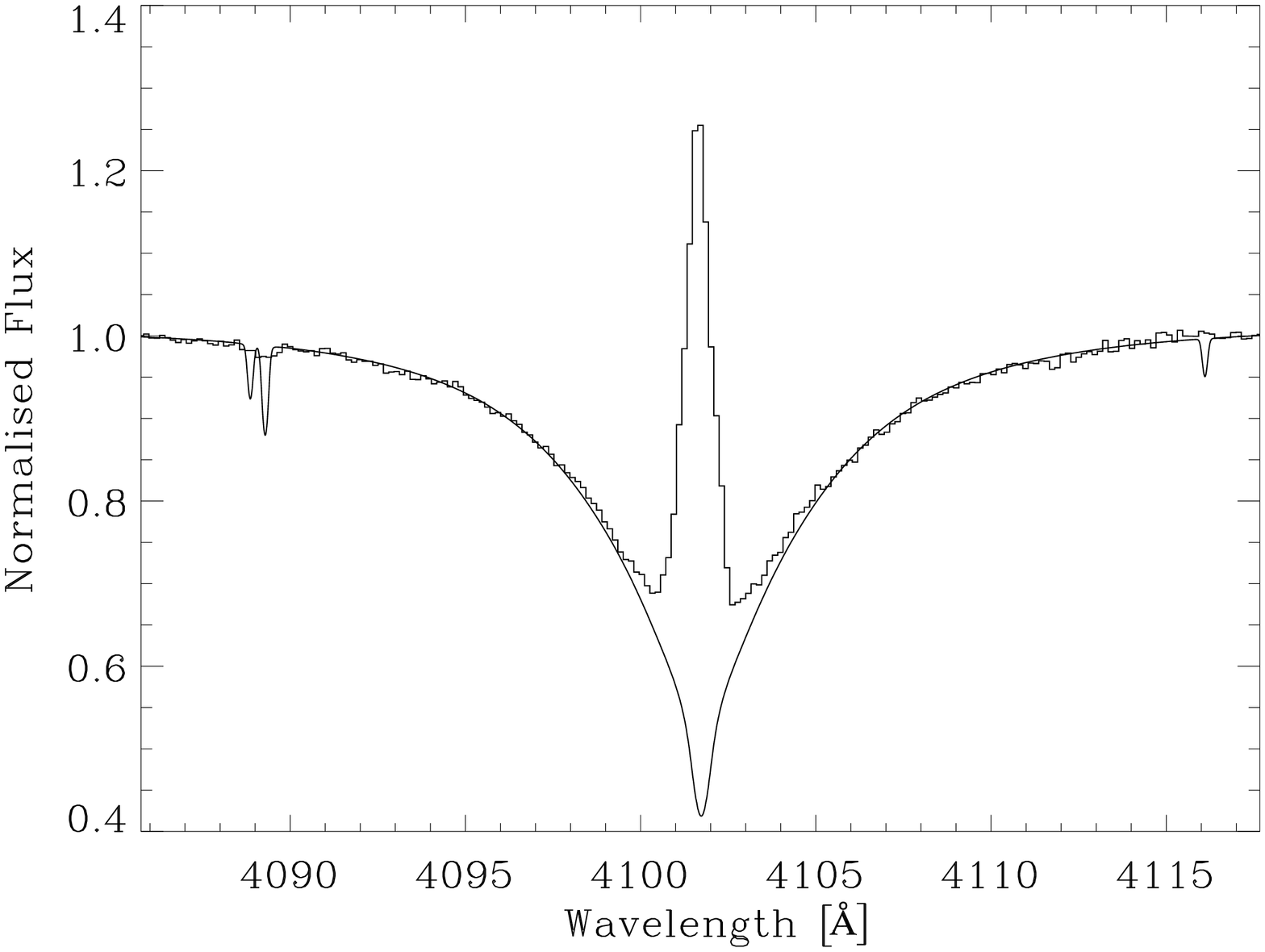,angle=90,width=0.37\textwidth} 
  \hspace*{-8mm} \psfig{file=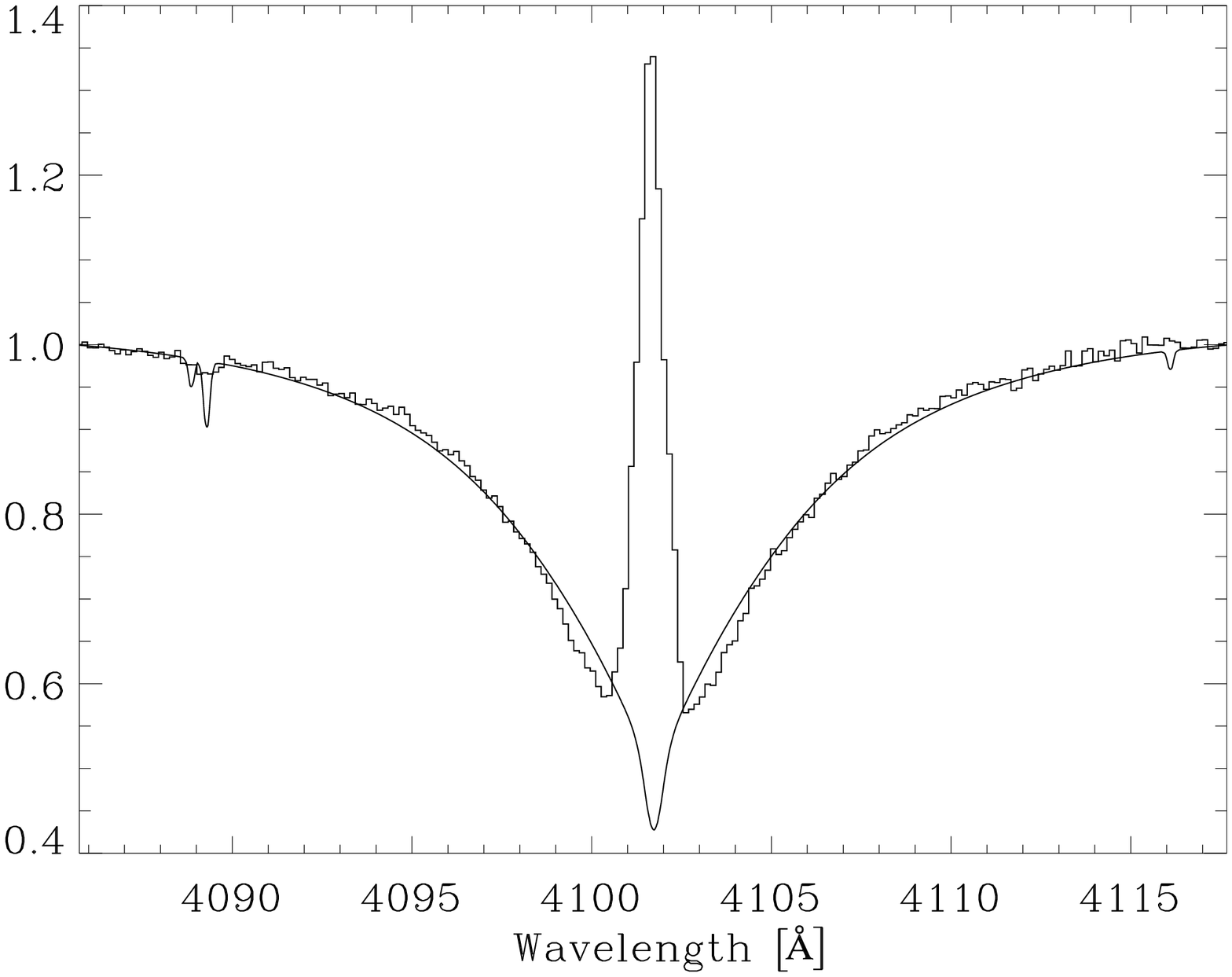,angle=90,width=0.37\textwidth} 
  \hspace*{-8mm} \psfig{file=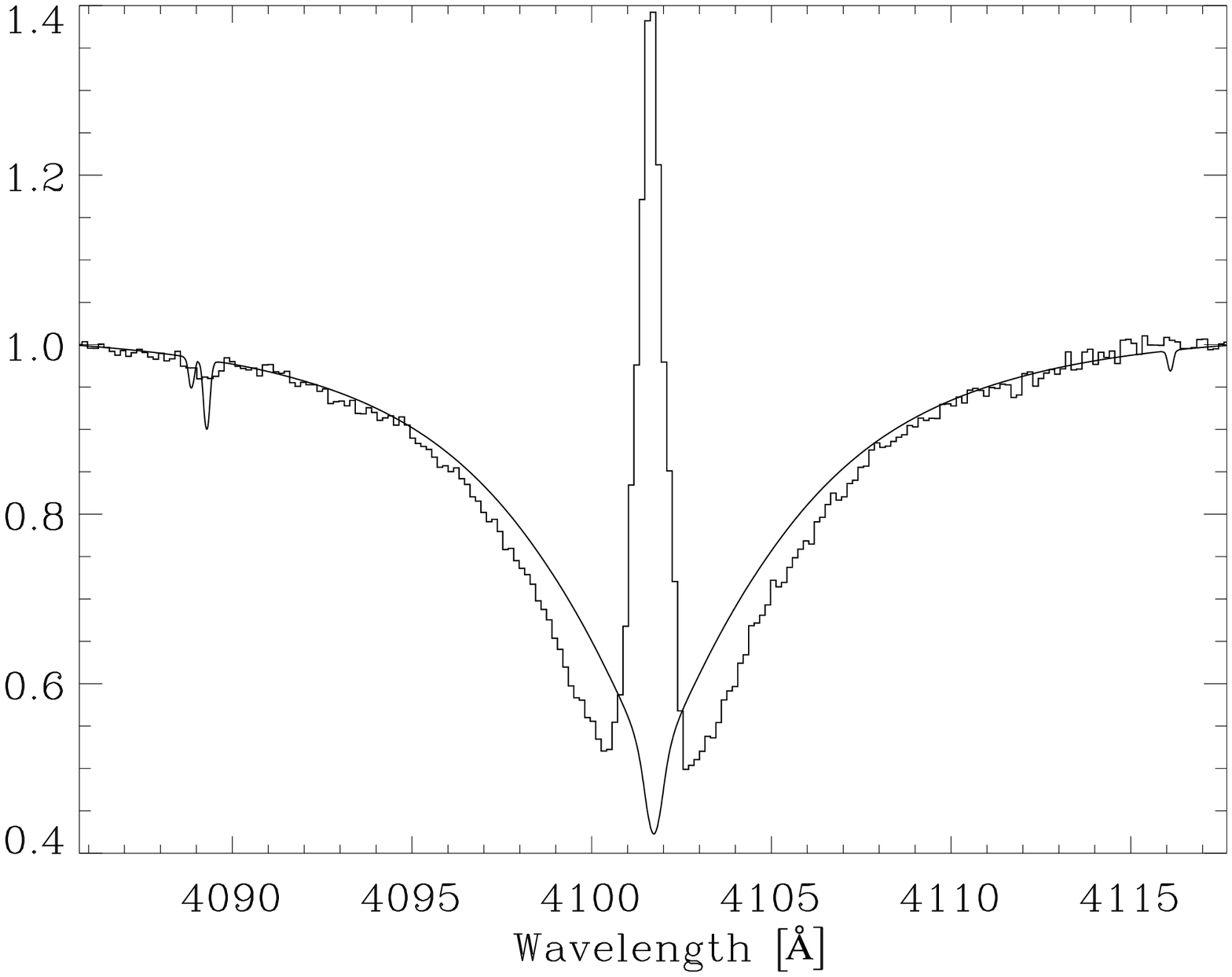,angle=90,width=0.37\textwidth}\\

  \caption{H\,$\delta$ lines with the zero, 25\% and 35\% continuum disk contribution
  and the fits by theoretical
  profiles modelled with the atmospheric parameters corresponding to the 
  low \teff\ case as presented
  in Table~\ref{tab:abs_abn}. Parameters for the 35\% model are \teff = 22\,750 K 
and log\,g = 3.95.}\label{fig:H_profiles}

 \end{figure*}

\subsection{Disk Contribution?}

Our program star
B\,12 is an extreme Be star which has {\it very} strong H\,$\alpha$ emission
from a face-on disk.  One can easily compute the degree of continuum veiling 
for a given absorption line, which is required to give a normal SMC abundance.
If the fractional disk contribution to
the continuum is defined as $x$, then the relationship between the measured
EWs, $w$, and its intrinsic EWs without a disk, $w_0$, will be given as $(1+x)
= (w_0/w)$. In practice, $w_0$ can be obtained by finding an EW that gives the
same abundance as AV\,304.  Such scaling factors for
each line are shown in Fig.~\ref{fig:scale}.  The error-bars are calculated by
assuming the measurement error of B\,12 is similar to that in AV304 and thus
only reflects the uncertainty in the observational measurement.   
No noticable dependence of $x$ by wavelength is seen in 
Fig.~\ref{fig:scale}, and the mean value of $x$ is $0.94
\pm 0.49$, implying  half of the total flux comes from the disk. 
The cooler model for B\,12 gives a lower mean value for $x$ of $0.60
\pm 0.32$, implying that 40\% of the total flux may be contributed
by the disk.

Could a disk contribute so much flux?  There is observational
evidence that the contribution of the disk can be quite
significant. For example, Telting et al.\ (\cite{Tel98}) have carried out a
long-term photometric monitoring of X\,Persei, a Be star  whose disk is
continuously being destroyed and reformed. This provides a good opportunity to
observe the star with and without the disk and thus to estimate the disk contribution. 
They showed that the $V$ magnitude can change by 0.6 to 0.7 magnitudes, a little
less in B, which implies that the disk may contribute up to 40\% of the total
flux.  This supports our assumption and leaves open the possibility that the
abundances of metal lines in B\,12 are in fact `normal' but appear to be
under-abundant due to continuum contamination.  

In a further refinement of this model we have undertaken a full model 
atmosphere analysis for
varying fractions of disk contribution, taking into account the change
in effective temperature and surface gravity implied by the disk-subtracted
and re-normalized spectrum. 
It is important to note that the derived 
surface gravity increases as the amount of continuum
subtraction increases, as does the effective temperature.  The increased
gravity leads to a decrease in the theoretical metal line strengths
and an even stronger increase in the abundance estimates
(summarised in Table~\ref{tab:abs_abn}).  Furthermore, the
Balmer lines enable us to estimate the maximum continuum contribution by
using the goodness of fit to their line cores, as shown in 
Fig.~\ref{fig:H_profiles}. Compared to the simple 
approximation of simply scaling line strengths, discussed above, this 
procedure reduces the fraction of veiling necessary to
obtain `normal' abundances, although as  Fig.~\ref{fig:H_profiles} shows, 
increasing the veiling beyond 25\% leads to unacceptable fits to the Balmer
lines.  In Table~\ref{tab:abs_abn} we list the abundances which 
result from assuming a disk contribution of 25\% for both estimates
of the effective temperature discussed above.  We can see that
this approach for B\,12 results in only moderate discrepancies with AV\,304,
particularly for the lower effective temperature model.


%

\subsection{Rapid rotation?}

Be stars are generally assumed to be fast-rotators, so it is important
to investigate whether or not our analysis, which is based upon non-rotating
models, is valid.   The answer to this question hinges on the issue
of the true rotational velocity of B\,12.   As has been shown by a number of
authors (Collins et al.\ \cite{Col91}, Howarth \& Smith \cite{How01}), 
the effect of rotation (through
von Zeipel gravity darkening) on line profiles of
OB stars only becomes sigificant for rotational velocities greater than
about 90\% of the critical value.  
Recently Chauville et al.\ (\cite{Cha01})
analysed a sample of 116 Be stars and concluded that most Be stars
rotate at 80\% of their critical velocity, with a very low dispersion
around this value. Such a rotational velocity is 
too low to have a serious impact on line strengths.
However this
result has been questioned by Townsend et al.\ (\cite{Tow04}) who show that stars rotating
very close to their critical velocities will have their rotational velocities
underestimated using the standard line-width techniques used by Chauville et al.
Essentially the line widths of B-type stars are insensitive to rotational
velocity for values above about 80\% of the critical velocity. As Townsend
et al.\ (\cite{Tow04}) point out, this raises the possibility that B-type stars rotate
much closer to their critical velocities than is generally believed. 
If this is true for B\,12 then it may well
be that the residual differences in abundances between B12 and AV304
(shown in Table~\ref{tab:abs_abn}), and the discrepancies in fitting the Balmer line
cores illustrated in Fig.~\ref{fig:H_profiles}, result from the neglect of these effects.

\subsection{Evolutionary Status}

It is very difficult to estimate the luminosity and hence the 
mass of B12 due to both the contribution of the disk
to the observed
magnitudes, and the unknown rotational velocity of the star (a rapidly
rotating pole-on star will appear brighter).
Nevertheless
a typical mass for a main-sequence B-star with the temperature
and surface gravity of B\,12 is $\sim$10 $M_\odot$, so it is not
unreasonable to assume that the mass of B\,12 lies near to this value.
Maeder \& Meynet (\cite{Mae01}) have published an evolutionary
track for a 9 $M_\odot$ star with SMC-like composition and an initial
rotational velocity of 300 \kms , well below the critical value which is above
450\kms.  As discussed by Lennon (\cite{Len03a}), this model implies a surface
nitrogen abundance of approximately 7.2 dex by the time it reaches the 
end of the main-sequence, and an equatorial velocity of approximately 
200 \kms.  Such a large nitrogen enhancement is clearly ruled out by the 
current observations. 
Furthermore, since B\,12 may be rotating even faster than
this model implies, the abundance discrepancy is probably being 
underestimated by this comparison.


Why does B\,12 have no nitrogen enrichment? It may well be the case that
rotational mixing is less efficient than current models predict,
in which case B\,12 can be used
to provide a strong constraint on the models. This seems unlikely
given the results of Lennon et al.\ (\cite{Len03b}) for other
B-type stars in the cluster, plus the other results for early-type
stars mentioned in the introduction.
Is it possible that B\,12 was a slow rotator which has
`recently' been spun-up?  Given
the large number of Be stars in NGC\,330, a spin-up mechanism such as mass-transfer
in a binary system,
needs to be quite common and co-ordinated in time, which seems unlikely.  

\begin{figure}\centering
 \begin{tabular}{l}
 \hspace*{-5mm}\psfig{file=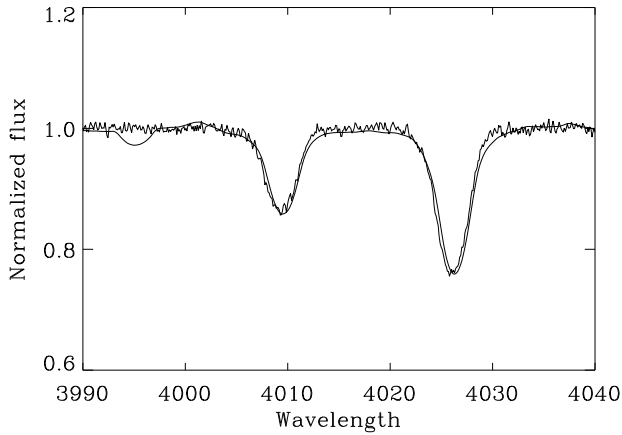,width=0.5\textwidth}\\
 \end{tabular} 
  \caption{Comparison of N\,{\sc ii} 3995 \AA\ region of the spectra of the Be star B\,17 with
the nitrogen rich B-type star B\,32 (smooth line), 
the latter having been convolved with a
rotational broadening fuction of 140 \kms. Note that the nitrogen line in
question is clearly absent in B\,17.
}\label{fig:b17}
 \end{figure}

Perhaps a more attractive solution lies in the recent work on
the influence of magnetic fields with rotation on stellar evolution
(see for example Maeder \& Meynet \cite{Mae04}).  
Their preliminary 15 $M_\odot$ model implies 
that magnetic fields may inhibit mixing of core-processed material
to the surface and further that they force models to rotate
as solid bodies while on the main-sequence. The latter point is
important since compared to rotating models without magnetic
fields, the models with magnetic fields are rotating faster at the end
of the main sequence.  Both of these characteristics, fast rotation and
a lack of evidence of mixing, are relevant to B\,12.  Is this typical of
Be stars?  This is a difficult question to answer since most
Be stars have such large projected rotational velocities that the upper limits on 
detection of the N\,{\sc ii} spectrum
at typical values of s/n provide no meaningful constraints.  However, we
note that Baade et al.\ (\cite{Baa02}) also obtained 25 UVES/VLT observations 
of B\,17, another Be star NGC\,330 which has only a moderately high $v$sin$i$
of 140 \kms.  We have therefore compared
its spectrum with that of the 
nitrogen rich B-star B\,32, which appears to be a good spectral analogue,
although we have convolved the spectrum of B\,32 with a rotational broadening
function of $v$sin$i$=140 \kms.  The nitrogen lines are clearly absent in B\,17
 (Figure~\ref{fig:b17}).
It is therefore
tempting to suggest that a characteristic of Be stars is that
they have magnetic fields which constrains them to rotate as solid bodies
and inhibits mixing.  Of course the existence of magnetic fields in Be stars
is a very attractive scenario from the point of view of explaining the
existence of the disk itself through magnetic 
compression (Porter \& Rivinius \cite{Por03}),
although some doubt has been cast on this mechanism by Owocki \& ud Doula 
(\cite{Owo03}).

\section{Conclusion}

We have investigated the spectrum of an extreme pole-on Be star NGC\,330-B\,12
in the
SMC and found it to be almost devoid of nitrogen lines. In fact its spectrum is
consistent with that of the normal SMC star AV\,304 and it is {\em inconsistent} 
with the predictions of stellar evolution models which include the effects of 
rotationally induced mixing.  We speculate that our findings lend support to
the idea that magnetic fields may be present in Be stars, inhibiting mixing,
and that the star is also rotating at a velocity close to its critical value
which in turn leads to some residual moderate abundance anomalies, an artifact 
of models not including the effect of von Zeipel gravity darkening.   
Clearly, the SMC's very low pristine nitrogen
abundance make it the ideal place in which to search for nitrogen enhancements
in fast rotators, and in Be stars.

\bigskip
\begin{acknowledgements}
JKL acknowledges financial support from PPARC (grant no G/O/2001/00173), 
and DJL acknowledges support through QUB's visiting fellow programme 
(grant no V/O/2000/00479).  We are indebted
to Thomas Rivinius and Dietrich Baade for providing us with their
UVES spectra of Be stars in NGC\,330, and Ian Hunter for making available his
analysis of AV\,304. 
\end{acknowledgements}

\end{document}